# Synthesis and self-assembly of lipid (DMPC)–conjugated gold nanoparticles

Subhasish Chatterjee[1,2,†], Markrete Krikorian[2], Harry D. Gafney[2], Bonnie Gersten[2]

[1] Department of Chemistry, Graduate Center, City University of New York, NY 10016, USA
[2] Department of Chemistry, Queens College, City University of New York, NY 11367, USA

**ABSTRACT**

Bio-conjugated nanomaterials play a promising role in the development of novel supramolecular structures, molecular machines, and biosensing devices. In this study, lipid-conjugated gold nanoparticles were synthesized and allowed to form a self-assembled monolayer structure. The nanoparticles were prepared by a phase transfer method, which involved the reduction of potassium tetrachloroaurate (III) by sodium citrate in an aqueous solution and the simultaneous transfer of the reduced species to an organic medium containing DMPC (1,2-dimyristoyl-sn-glycero-3-phosphocholine). The gold nanoparticles were characterized using UV-Vis spectroscopy and dynamic light scattering (DLS) particle-size analysis. In addition, the resulting nanoparticles were examined using transmission electron microscopy (TEM). The Langmuir-Blodgett (LB) technique was used to assemble the DMPC-capped nanoparticles onto a water subphase at room temperature. The measurement of the compression isotherm confirmed the assemblage of lipid capped gold nanoparticles. This method of synthesis of ordered structures utilizing molecular interactions of lipids will be useful in developing novel metamaterials and nanocircuits.

---

[†] schatterjee@gc.cuny.edu

# INTRODUCTION

The assembly of bio-conjugated colloidal nanomaterials has attracted extensive interest due to their tunable physiochemical properties that can be controlled as a function of their size, shape, aggregation state and local environment [1, 2]. Hybrid organic-inorganic nanostructures have versatile technological potential for the development of integrated circuits [2], molecular machines [3], and biosensors [4]. Vigorous attempts have made to obtain assembled structures of metallic [5] and semiconductor [6, 7] nanostructures (nanoparticles, nanorods) [8, 9] using various molecular systems, such as polyelectrolytes [1, 3], surfactants [4], and long-chain hydrocarbons [10]. The ability to precisely control the overall size of an assembled structure and the separation distance between nanoparticulate remains a significant technical challenge [1-4, 11], yet the uniformity in an ordered structure controls the physical properties of the entire architecture [1-3]. Additionally, the stability of nanoparticles in their assembled forms is imperative for the development of an integrated structure [1-4]. The interaction among capping reagents assists to maintain the stability of nanoparticles and the uniformity of their size [11]. Thus, the capping molecules play a crucial role in the formation of a stable assembly of nanoparticles as well as limiting the overall dimension of the assembled architecture [1, 4]. Moreover, it is crucial to retain the native properties of biomolecular capping reagents after conjugating with nanomaterials, which play a critical role in establishing an ordered structure [1, 3, 11].

Phospholipids, integral components of cell membranes, are composed of polar head groups and nonpolar hydrocarbon tails [12]. Because of the amphiphilic nature of lipids, it is possible to modulate biomolecular interactions by altering the hydrophobic and hydrophilic components of their molecular frame. Hence, the use of phospholipids as capping reagents can facilitate the syntheses of a wide variety of biocompatible nanomaterials [13]. Many variations in the size and shape of the assembled nanostructures can be achieved by controlling the separation distance between nanoparticulates with atomic precision using selective molecular interactions of lipids [14, 15]. In addition, lipids can affect the dielectric properties of metallic nanoparticles, providing a way to modulate the optical properties of an integrated architecture and serve as optical sensors [4]. Through altering the nature of the polar head groups, the interaction between the lipid and nanoparticle can be adjusted [13]. The size and number of hydrocarbon tails can control dimensions and kinetics to form an assembly of lipid-capped nanomaterials. Additionally, various types of lipids can be mixed [15] to alter the separation distance between nanoparticles. Owing to the lipid's amphiphilic characteristics, lipid-capped nanoparticles can be fabricated on substrates of various surface polarities [12]. Accordingly, a specific directionality in the growth of an assembled structure can be achieved to develop novel supramolecular architectures. Since the study of the phospholipids' assembly is used to model biological membranes [16], an investigation of the hybrid structures of metallic nanoparticles and phospholipids has potential for the development of new biosensing devices and drug delivery methods across cellular membranes.

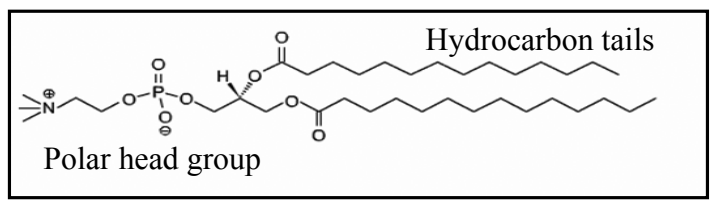

**Figure 1:** DMPC (1, 2-dimyristoyl-sn-glycero-3-phosphocholine)

Here we report the synthesis of gold nanoparticles directly conjugated to DMPC (1, 2-dimyristoyl-sn-glycero-3-phosphocholine), a phospholipid with two hydrophobic tails of 14 carbons each (Figure 1). The size and shape of the gold nanoparticles were estimated in a solution phase and on a solid support using dynamic light scattering (DLS) and transmission electron microscopy (TEM), respectively. The Langmuir-Blodgett (LB) method, commonly applied to assemble long-chain amphiphilies on a liquid subphase or a solid substrate forming lamellar structures of lipids [16, 17], was utilized to form an assembled monolayer of lipid-capped nanoparticles. Comparative analyses between assemblies of the DMPC capped gold nanoparticles and unmodified DMPC molecules were performed to understand the effect of the direct conjugation of the gold nanoparticle on the lipid. By extending this method, the fabrication of nanoparticles films on a substrate can be achieved to develop solid-state devices [12].

**EXPERIMENTS**

**Materials**

DMPC (1, 2-Dimyristoyl-sn-glycero-3-phosphocholine, Avanti Polar Lipids), toluene (HPLC grade, Sigma-Aldrich), potassium tetrachloroaurate (III) ($KAuCl_4$) (98%, Sigma-Aldrich), sodium citrate ($Na_3C_6H_5O_7$ X $2H_2O$) (99.0%, VWR), Millipore water.

**Synthesis of DMPC-capped gold nanoparticles**

The principle of phase transfer synthesis, generally applied to synthesize surfactants and amphiphiles capped nanostructures [2, 11, 13], was used to prepare the DMPC-capped gold nanoparticles in toluene. A solution of DMPC (0.25 mg/ml) in toluene was prepared in a 30 ml glass vial and stirred for 15 mins to ensure complete dissolution of the lipid in toluene. An aqueous solution of potassium tetrachloroaurate (III) (1.0 mg/ml) was prepared; 2 ml of the solution was added to 5 ml of the lipid solution and the mixture was stirred for 15 mins. While the mixture was being stirred, 1 ml of an aqueous solution of sodium citrate (10 mg/ml) was added dropwise. The reaction mixture was stirred for 8 hrs. Towards the end of the stirring, the organic phase turned into a clear deep red solution and the aqueous phase beneath it was an opaque gray milky color. The organic layer was removed with a glass syringe.

### Characterization of the gold nanoparticles

An Ocean Optics HR2000 spectrometer was used to study the UV-Vis absorption of gold nanoparticles in toluene that were placed into a 3.5 ml quartz cuvette of 1 cm path-length.
The hydrodynamic diameter and polydispersity of the gold nanoparticles in toluene were obtained at room temperature using a BIC 90plus DLS particle-size analyzer. The size and shape of the lipid-capped gold nanoparticles were investigated using a JEOL EX-1200 TEM after the nanoparticles were drop coated on a carbon coated copper TEM grid (300 mesh with formvar coating on one side, SPI supplies, Structure probe, Inc).

### Assembly of DMPC and DMPC capped gold nanoparticles using Langmuir-Blodgett Apparatus

A Kibron Inc. (KBN315) Microtrough X was utilized to allow the self-assembly of the DMPC-capped gold nanoparticles onto a water subphase at room temperature. 20 µl of the nanoparticles solution (DMPC concentration 0.25 mg/ml) was deposited onto a water surface using a Hamilton manual gas-tight microsyringe. After allowing 20 min to evaporate the organic solvent, the nanoparticles were compressed at the rate of 4.68 Å$^2$/ (molecule)/min. The compression isotherm was measured to monitor the self-assembly of the DMPC-conjugated gold nanoparticles. The same method was followed to assemble unmodified DMPC.

**RESULTS AND DISCUSSION**

The optical absorption spectrum of the gold nanoparticles exhibited a peak at 525.73 nm, supporting the formation of stable nanoparticles in toluene (Figure 2) [4,13].

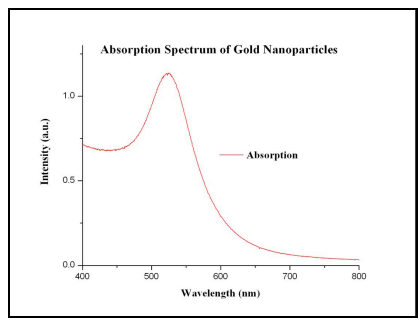

**Figure 2**: UV-Vis absorption spectrum of gold nanoparticles in toluene.

DLS utilizes temporal variations of fluctuations of the scattered light to measure an average hydrodynamic diameter of particles suspended in a liquid medium [18]. The results of dynamic light scattering showed the size of the lipid coated gold nanoparticles to be 31.5 nm with a polydispersity of 0.205 (Figure 3).

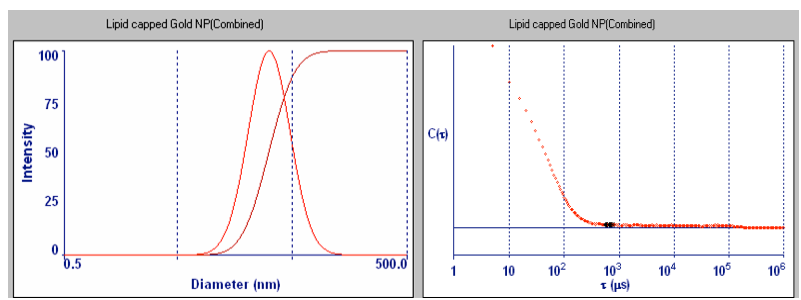

**Figure 3:** DLS analysis of gold nanoparticles from a lognormal plot of intensity vs. diameter of the nanoparticles. Fluctuations in the time intensity of the scattered light were processed by computing the autocorrelation function, C ($\tau$), $\tau$ = decay time.

TEM showed the lipid capped gold nanoparticles maintained a spherical shape, with diameters ranging from 15 to 25 nm (Figure 4).

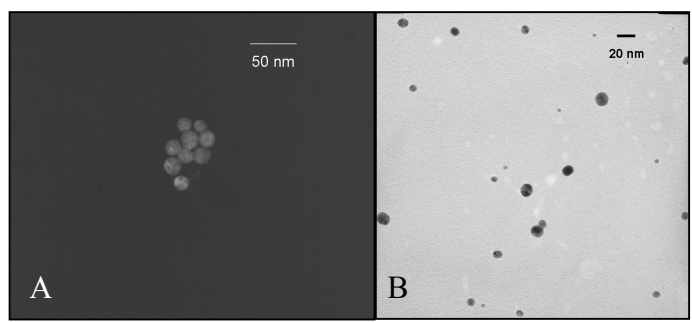

**Figure 4**: TEM images of gold nanoparticles at a magnification of 120k (A) and 200k (B).

The average hydrodynamic diameter of the gold nanoparticles obtained from DLS relies on intensity-weighting of the scattered lights, which could be biased towards larger particles due to their dominant effect on the light scattering. Although TEM imaging directly elucidates the geometric size and shape of the gold nanoparticle, it probes the colloidal nanoparticle out of the solution phase and its native environment.

The characteristic surface plasmon absorption (Figure 2) confirms the formation of the nanoscopic gold particulates in the solution phase [2, 4]. The spherical shape of the nanoparticles and their well-separated disposition in the TEM images (Figure 4) can conceivably be ascribed to the lipid capping molecules of the gold nanoparticles. Considering the structural similarities between lipids and surfactants [12, 13] and the proposed nature of the interaction of surfactants with metallic nanoparticles [2, 11], it could be suggested that the polar head group of DMPC stabilizes the nanoparticles by forming non-covalent bonds. The hydrocarbon tails of the attached lipids extend beyond the surface of the nanoparticle, assist in maintaining a stable solution in non-polar toluene by providing a hydrophobic shield preventing nanoparticle aggregation.

The compression isotherms (Figure 5A & 5B) indicate the formation of a monolayer for both unmodified DMPC and the lipid-capped gold nanoparticles onto a water surface.

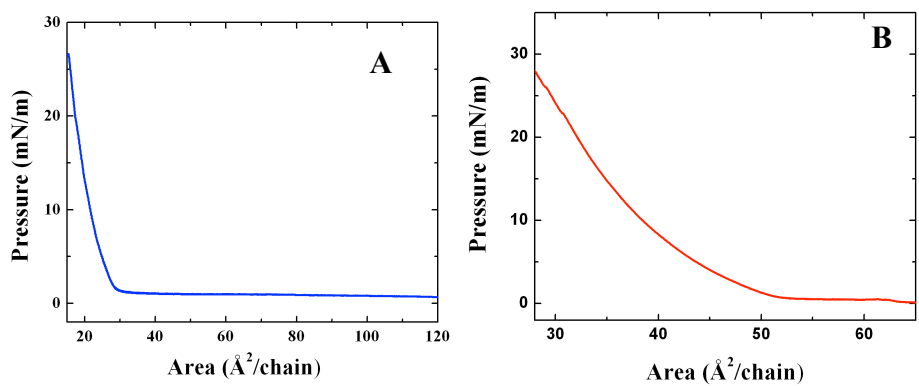

**Figure 5:** Compression isotherm ($\pi$–A) curves of DMPC (**A**) and DMPC-conjugated gold nanoparticles (**B**).

In general, theoretical modeling of the formation of a monolayer of lipids onto a water surface using the LB method shows a first-order phase transition [16] and the pressure-change in the compression isotherm indicative of formation of a new phase (Figure 5A). Initially, lipids form as a "gas-like" dispersed phase on a water subphase, keeping their hydrophobic tails away from the aqueous phase [16, 19]. With the application of a directed force by movement of the LB barriers, the lipids form an ordered assembly on top of the polar aqueous phase because of the strong hydrophobic interaction of the lipid's hydrocarbon tails [16, 17, 19]. The lipids conjugated to the nanoparticles exhibited a similar behavior, which supports the inference that the associated lipids retained their ability to form an assembled structure. Due to the association with the gold nanoparticles, the modified DMPC experienced a localized ordering, causing a shift of the rising point of the pressure (kink at 51.008 Å$^2$/chain) at the compression isotherm curve (Figure 5B).

**CONCLUSIONS**

In this study, the ability to synthesize DMPC-capped gold nanoparticles and to form an assembly of the phospholipid-conjugated gold nanoparticles onto an aqueous subphase was tested. It was found that the direct conjugation of DMPC stabilized the nanoparticles, and conversely bonding to nanoparticles controlled the conformation of the lipids. Thus, lipid molecules conjugated with nanoparticles exhibited a relatively different compression isotherm compared to the original lipid molecules assembled alone. The change in the pressure of the compression isotherm signals the assembly of the lipid molecules. Accordingly, it should be possible to develop nanoparticles capped directly with lipids as self-assembled and controllable structures.

ACKNOWLEDGMENTS

The authors thank Prof. Ruth Stark, Prof. Vinod Menon, Dr. Nikesh Valappil, Dr. Baohe Chang, Dr. Areti Tsiola, and Ms. Miriam Ginzberg, for their assistance. This work was supported by grants from the PSC-CUNY Research Award program, Graduate Center Dissertation Fellowship, and by the James D. Watson NYSTAR Award.